\documentclass[a4paper,12pt]{article}
\usepackage{amssymb,graphicx,epsfig}
\usepackage{lscape,graphics,amsmath,geometry}
\usepackage{latexsym}
\usepackage{axodraw}
\usepackage[english]{babel}
\usepackage{graphics}
\newcommand{\wtb}{$Wtb~$}
\newcommand{\dzero}{$D0~$}
\newcommand{\cms}{$CMS~$}

\begin{document}

\vspace*{0.7cm}

\begin{center}
{\Large\sc {\bf Modeling of processes with anomalous interactions by means of auxiliary fields}}
\vspace*{0.7cm}

{\sc E. Boos}$^{1}$, {\sc V. Bunichev}$^{1}$,  
{\sc L. Dudko}$^{1}$ and {\sc  M. Perfilov}$^{1}$
\begin{small}
\vspace*{0.9cm}

$^1$  Skobeltsyn Institute of Nuclear Physics, MSU,
119992 Moscow, Russia. 
\vspace{3mm}
\end{small}
\end{center}
\vspace*{0.7cm}

\begin{abstract}
The paper presents a method of modeling events with anomalous fermion-boson couplings by means of an auxiliary vector fields in addition to the SM gauge filed. The method allows to simulate anomalous gauge couplings in different approaches, keeping only the linear order or higher order contributions of the anomalous couplings. 
The correctness of proposed method is demonstrated by modeling the single top quark production  processes with anomalous \wtb~couplings. 
\end{abstract}

\section{Introduction}
\label{Introduction}

The search for physics beyond the Standard Model (BSM) is one of the main 
tasks of the energy-frontier colliders specially in a view of new 
Higgs-like boson discovery at the LHC.   

In general, possible manifestations of ``New Physics'' 
depend on relations between the characteristic 
collision energy $(E_{\rm  collisions})$, which is typically a few TeV at the LHC, and thresholds of 
possible new states. If the collision energy is greater than the 
production threshold $(E_{\rm  collisions} > E_{threshold})$~new states 
predicted in many BSM scenarios can be produced directly. If the 
collision energy is smaller than the production threshold $(E_{\rm 
collisions} < E_{threshold})$ new states can not be produced directly and 
``New Physics'' may manifest as deviations from SM predictions in production cross sections 
and/or kinematic distributions due to contributions of possible anomalous couplings originating
from higher dimentional gauge invariant operators or  
interference terms of new resonances with the SM contribution below the resonance thresholds. 

A standard way to compute and model physics effects below thresholds  is the 
effective Lagrangian approach in which new physics contributions are 
encoded in a serie of higher dimensional operators preserving the Standard 
Model gauge invariance \cite{Peccei:1989kr}, \cite{Buchmuller:1985jz}. 
It is straightforward to derive Feynman 
rules from the effective operators, add them to the SM vertices, insert 
in some computer code and perform the necessary  computations and event 
generation. However, in practice there are a number  mainly technical but 
time consuming difficulties. If unstable particles, such as EW gauge W 
and Z bosons, the Higgs boson, the top-quark etc., participate in the 
process under study the anomalous couplings may contribute to both 
production and decay, as well as to the decay width of the unstable  
particle. The corresponding matrix element is then a ratio of polinomials  
of anomalous parameters affecting not only the 
production rate and decay branching fraction but also spin 
correlations and corresponding distributions. Even when the leading 
$1/{\Lambda}^2$~effect is analyzed, motivated by the consistency of the 
effective operator expansion \cite{Zhang:2010dr}, it is quite involved 
to combine all the necessary terms coming from different places of the matrix 
element. Usually there are several operators contributing to the same vertex,
and the problem the large number of event samples one should 
generate in order to get useful information about all  anomalous parameters. 
This is specially problematic 
 in preparing Monte-Carlo event samples for experimental 
searches because of the computer time and memory consuming procedure of event
 propagation through detector simulation programs.

In this paper we describe a simple practical method to reduce these difficulties significantly. 
The method is based on an implementation of additional auxiliary vector fields. It is directly applicable 
for automatic computer codes.  We demonstrate the method effectiveness 
by modeling the
single top production and decay process involving  anomalous Wtb top quark couplings 
using the CompHEP \cite{Boos:2009un} code.  

\section{Main idea of the method}
\label{Idea}
The presence of anomalous parts due to contributions of some effective 
operators can be written in the form of additional terms to the SM gauge-fermion vertex: 
\begin{equation}
\Gamma_{\mu} = \Gamma_{\mu}^{SM} + \Gamma_{\mu}^{NP_1} + \Gamma_{\mu}^{NP_2} + ...,
 \label{vertex}
\end{equation}
where the first term is the SM vertex of the SM gauge boson ($V_{\mu}$) and the 
SM fermions, ($f$), and the 
other terms are the anomalous (New Physics) parts of the vertex with 
all the allowed Lorents structures such
as $\gamma_{\mu}$, $\gamma_{\mu} \gamma_5$, $\sigma_{\mu\nu} k^{\nu}$ etc.

The idea is that for every part of the anomalous vertex  
$\Gamma_{\mu}^{NP_1}$, $\Gamma_{\mu}^{NP_2}$ etc.
one adds to the SM a new auxiliary vector bosons $V_{1~aux}^{\mu}$, $V_{2~aux}^{\mu}$ etc. 
with the same mass and all the other couplings
as for the SM gauge boson $V^{\mu}$, but with the coupling to the fermion $f$  
being $\Gamma_{\mu}^{NP_1}$, $\Gamma_{\mu}^{NP_2}$ etc.,
respectively.

Now the model contains the SM boson $V^{\mu}$ with the SM model couplings and 
a number of auxiliary vector bosons $V_{1~aux}^{\mu}$, $V_{2~aux}^{\mu}$ etc.
 with the vertex to the fermion $f$ being 
non-standard $\Gamma_{\mu}^{NP_1}$, $\Gamma_{\mu}^{NP_2}$ etc. and with 
all the other SM couplings. The total width of the fermion $f$ should include all anomalous contributions.

For a process involving $V^{\mu}$ in the intermediate state 
there will be several contributing Feynman 
diagrams in addition the SM diagram involving new bosons $V_{1~aux}^{\mu}$, $V_{2~aux}^{\mu}$ 
etc.. 
The sum of all these diagrams 
is exactly the same as for the case when we simply change the SM vertex $\Gamma_{\mu}^{SM}$ 
to the new vertex $\Gamma_{\mu}$~(Eq. \ref{vertex}).

The virtue of this approach is that one can easily switch on or off 
contributions of any new auxiliary boson, keep and compute any interference terms
with those  of the SM,
compute, if needed, only linear terms of new contributions, and so on. In the next section, it is shown how such a simple method works 
in a real example.

\section{Anomalous \wtb~couplings in the single top quark processes.}
\label{sec:anom_wtb_in_stop_processes}

In this Section, we demonstrate how the introduced earlier method works for the practical example related to  anomalous \wtb~couplings searches in the single top quark production processes. 
The anomalous terms in Wtb vertex allowed by the Lorentz invariance are parametrized 
by the following effective Lagrangian as was proposed in~\cite{Kane:1991bg}: 

\begin{eqnarray}
\label{anom_wtb_eq_lagrangian}
 {\cal L} = - \frac{g}{\sqrt{2}} \overline{b}{\gamma}^{\mu} \big( f^L_V P_L + f^R_V P_R \big) t W^{-}_{\mu} - \frac{g}{\sqrt{2}} \overline{b} ~\frac{ i{\sigma}^{\mu\nu} q_{\nu} }{M_W} \big( f^L_T P_L + f^R_T P_R \big)t W^{-}_{\mu\nu} \nonumber \\
 + h.c. 
\end{eqnarray}

Here $M_W$ is the W-boson mass, $P_{L} = (1 - \gamma_5)/2 $ is 
the left-handed projection operator, $P_{R} = (1 + \gamma_5)/2 $ is 
the right-handed projection operator,  
$W^{\pm}_{\mu\nu} = D_{\mu}W^{\pm}_{\nu} - D_{\nu}W^{\pm}_{\mu}$, 
$ D_{\mu} = \partial_{\mu} - ieA_{\mu} $, 
$ {\sigma}^{\mu\nu} = i/2 \big[ \gamma_{\mu}, \gamma_{\nu} \big] $,
and $q_{\nu}$ is the four-momentum of $W$-boson. Parameters 
 $f^L_{V(T)}$ and  $f^R_{V(T)}$ are the dimensionless coefficients 
that parametrize strengths of the left (vector and tensor) and 
the right (vector and tensor) structures in the Lagrangian. 
In the SM all fermions interact through the left-handed currents  and all constants are equal 
to zero, except $f^L_V=V_{tb}$ (CKM-matrix element).

One should note that  all the terms in the Lagrangian (\ref{anom_wtb_eq_lagrangian}) 
come from various effective gauge invariant dimension-6 operators 
as given in~\cite{Boos:1999ca, AguilarSaavedra:2008gt}. 
Therefore, the natural size of the strength parameters are 
of the order of $(v/{\Lambda})^2$) where $v$ is the Higgs vacuum
 expectation value and $\Lambda$ is a scale of "New Physics". 

 Study of the anomalous top quark \wtb~interactions was a subject of intensive recent experimantal (\cite{Aad:2015yem},\cite{Khachatryan:2014vma},\cite{CMS:2014ffa}) and theoretical ( \cite{AguilarSaavedra:2011ct}, \cite{Fabbrichesi:2014wva}, \cite{Bernardo:2014vha}, \cite{Aguilar-Saavedra:2015yza}, \cite{Hioki:2015env}) studies. 
 
Squared matrix element for the single top production process  has a quadratic dependence on 
the anomalous constants of the following form (assuming a massless $b$-quark) \cite{Boos:2010zzb}:

\begin{eqnarray}
\label{xsection}
|M|^{2}_{pp \rightarrow t \overline{b}} \sim A \cdot (f^L_V)^2 + B \cdot (f^R_V)^2 
+ C \cdot (f^L_V f^R_T) + D \cdot (f^R_V f^L_T) \nonumber \\
+ E \cdot (f^L_T)^2 + G \cdot (f^R_T)^2,
\end{eqnarray}
where $A, B, C, D, E, G$ are functions of particle momenta. As one can see there are no $(f^L_V f^R_V)$ and $(f^L_T f^R_T)$ cross terms in Eq.~\ref{xsection}. 
The contributions of the left and right parts to  matrix element squared are different in common case ($A$ is not equal to $B$, and so on) because the SM vertex of W boson with the light quarks has the left (V-A) structure. 

In the same manner as it was done in the experimental searches for the anomalous \wtb~couplings~\cite{Abazov:2009ky,Abazov:2011pm,CMS:2014ffa}) we also considered three scenarios with always the left-vector coupling pairing with one of  the other couplings ($(f^L_V, f^R_V)$, $(f^L_V, f^L_T)$, $(f^L_V, f^R_T)$ scenarios).

\begin{figure*}[!!h!]
\begin{center}
 
\begin{minipage}[t]{.45\linewidth}
\centering
\includegraphics[width=60mm,height=50mm]{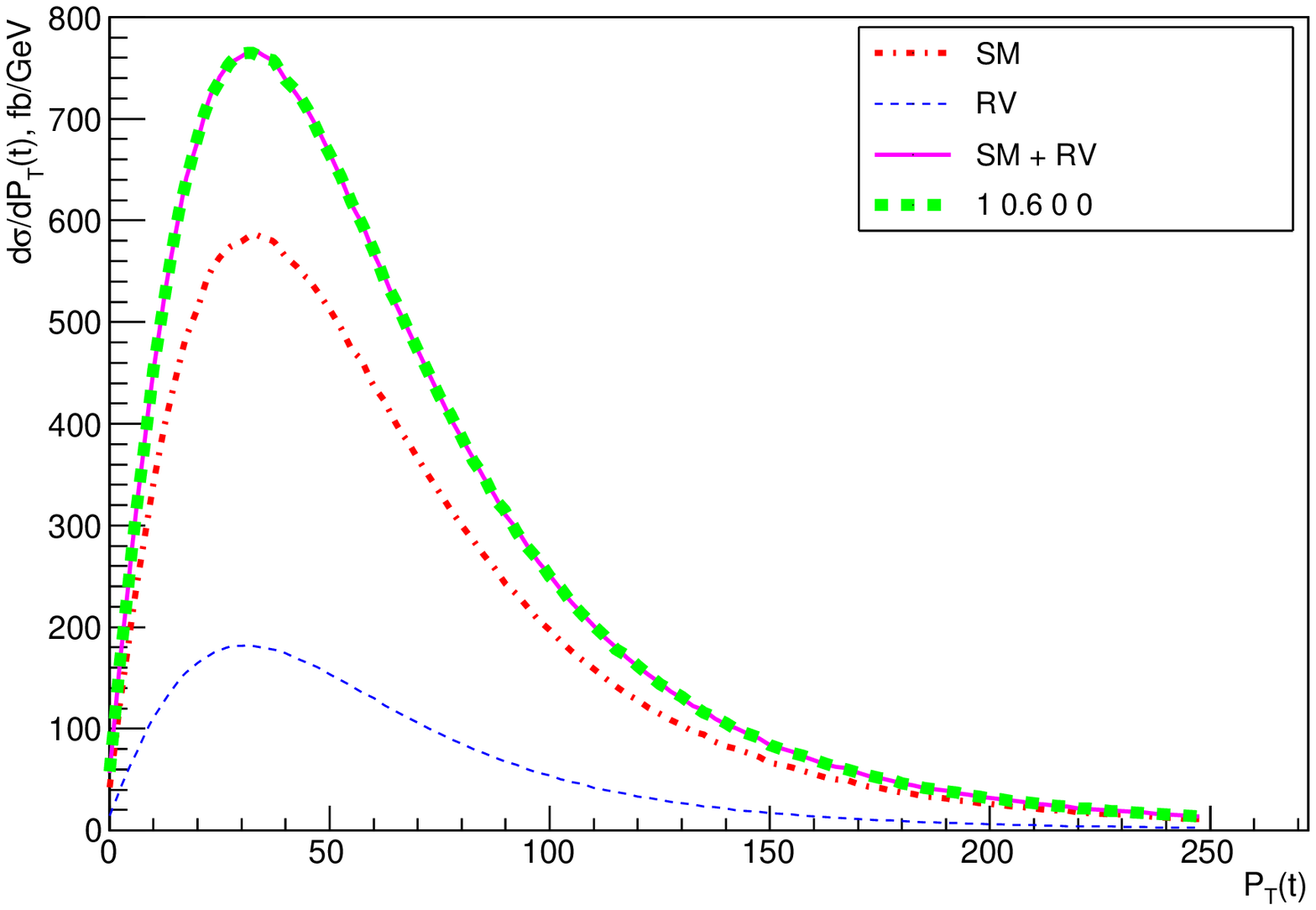}
\caption{\label{pic1} \footnotesize The transverse momentum distribution of the top-quark for the process $pp \rightarrow t q $ (t-channel) for the $(f^L_V, f^R_V)$ scenario. \label{prod_vs_prod_and_decay_a}}
\end{minipage}
\hfill
\begin{minipage}[t]{.45\linewidth}
\centering
\includegraphics[width=60mm,height=50mm]{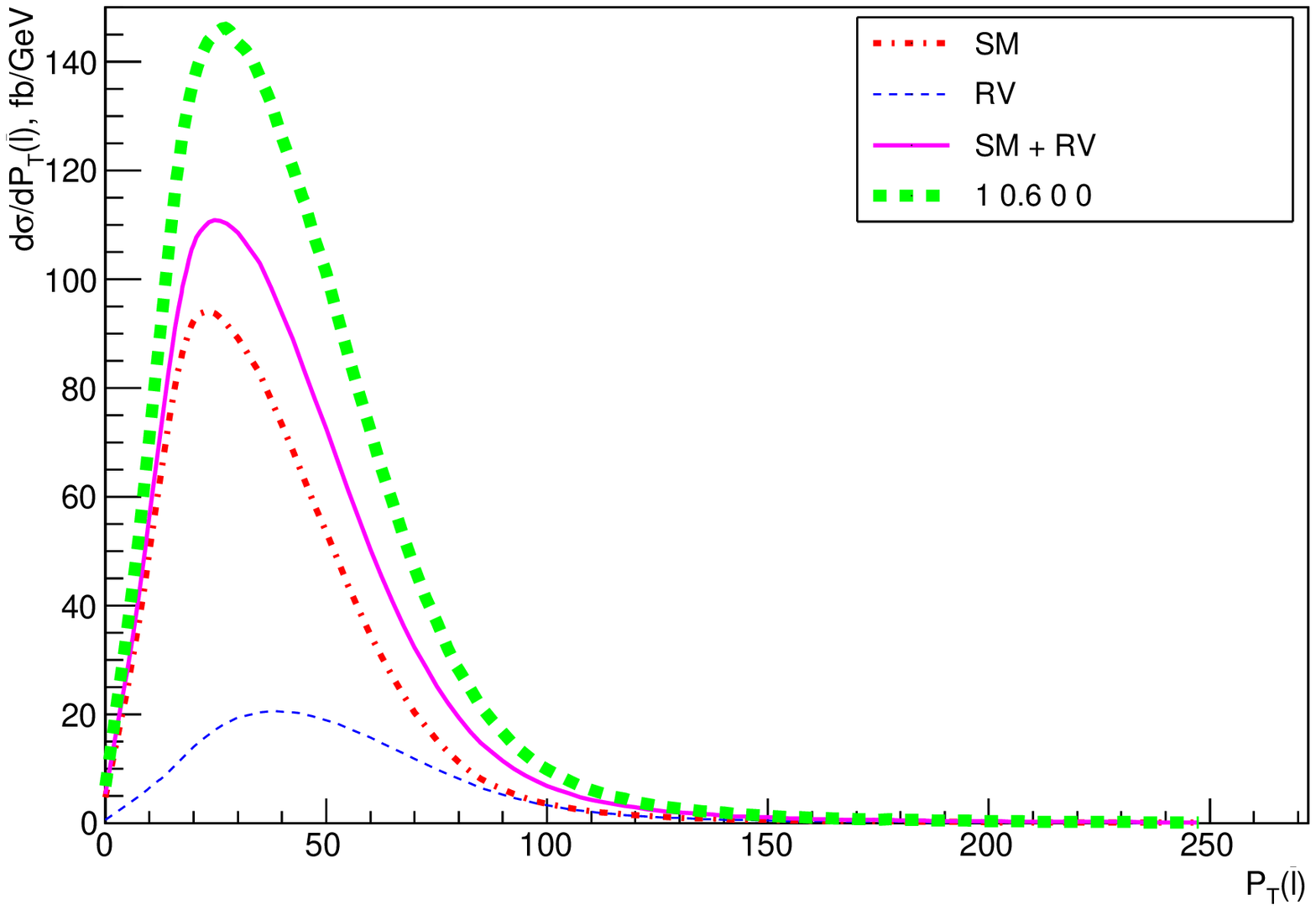}
\caption{ \label{pic2} \footnotesize  The transverse momentum distribution of the lepton from top quark decay for the process $pp \rightarrow t (\nu_l,\bar {l}, b)q$ for the $(f^L_V, f^R_V)$ scenario. \label{prod_vs_prod_and_decay_b}}
\end{minipage}

\end{center}
\end{figure*}

The first scenario has $(f^L_V, f^R_V)$ non-zero couplings in \wtb~vertex ($(f^L_V, f^R_V)$ scenario). The events of the top quark production processes without top quark decay can be simulated with only two sets of events in this scenario. 
The first set of Monte-Carlo (MC) simulated events corresponds to kinematic term $A$ 
in Eq.~\ref{xsection} and can be obtained with the fixed values 
of the anomalous parameters: $f^L_V=1, f^R_V=f^L_T=f^R_T=0$. In this sample,
 only events with left-handed interaction in the \wtb~vertex are simulated;  the notation "1000\_prod"  for this set of events is used. The second set of events corresponds to the kinematic term $B$ in Eq.~\ref{xsection} 
and can be simulated with the 
$f^L_V=0, f^R_V=1$ coupling values; this set of events corresponds to the only right-handed 
interaction in the \wtb~vertex. 
The notation for such sample is ''0100\_prod''.
As follows from Eq.~\ref{xsection} the kinematic distributions of a top quark production processes with all possible values of the $(f^L_V, f^R_V)$ anomalous couplings can be reproduced 
by the sum of corresponding distributions following from the ''1000\_prod'' and ''0100\_prod'' sets of events multiplied by the squared value of the corresponding coupling. This is demonstrated in  Fig.~\ref{prod_vs_prod_and_decay_a} for the distribution of transverse momentum of the top quark for some concrete values of the couplings are equal to $f^L_V=1, f^R_V=0.6$.   The notations for ${f^L_V}^2\cdot(1000\_prod)$ is ''SM''' and ${f^R_V}^2\cdot(0100\_prod) $ is ''RV'' at the plot. One can see that the sum of the curves "SM''' and ''RV' is precisely coinsides with the curve ''1 0.6 0 0'' which shows the distribution of the transverse momenta of the top quark with $f^L_V=1, f^R_V=0.6$ values of the couplings in the top quark \wtb~production vertex. 
%The 
%=========================================================================
%=============================================================================

However,  two sets of events "1000\_prod"  "0100\_prod" are not enough if one considers the top quark production with subsequent decay (the usual picture for experimental analyses). As one can see from Fig.~\ref{prod_vs_prod_and_decay_b} the distribution of the transverse momentum of the lepton from the top quark decay cannot be reproduced with a simple sum of the distributions following from "SM" and "RV" sets of events in the same way as it was done for the case with the anomalous couplings in the top quark production only (Fig.~\ref{prod_vs_prod_and_decay_a}).
Therefore, one needs to extend the minimal number of the sets of events in order to correctly represent the event kinematics and cross section rate for the case where the anomalous couplings are both in the production and in the decay vertices. 

\subsection{Production and decay of top quarks with anomalous couplings in $(f^L_V, f^R_V)$ scenario}
\label{subsec:lvrv}

The general squared matrix element structure of the single top quark production and subsequent decay with only $f^L_V$ and $f^R_V$ couplings are in the \wtb~vertex in the narrow width approximation is the following:

\begin{equation}
\label{me_sq_pr_decay}
  |M|^{2}_{tot}  \sim  \big ( (f^L_V)^2 A_p + (f^R_V)^2 B_p \big ) \frac {\big ( (f^L_V)^2 A_d + (f^R_V)^2 B_d \big )}{w_{tot}~(f^L_V,f^R_V)}
\end{equation}

where $A_p$ and $B_p$ ($A_d$ and $B_d$) are the functions of momenta in 
the production (in the decay) of the top quark. The total top quark width $w_{tot}~(f^L_V,f^R_V)$ 
can be obtained from \cite{Mohammadi Najafabadi:2006um} where the top quark total width in presence of all four anomalous couplings is presented.

After a multiplication we get the following expression:

\begin{multline}
\label{me_sq_pr_decay_1}
 |M|^{2}_{tot} \sim \frac {1}{w_{tot}~(f^L_V,f^R_V)} \big ( (f^L_V)^4 A_p A_d + (f^L_V)^2 (f^R_V)^2 A_p B_d + (f^L_V)^2 (f^R_V)^2 A_b B_p + (f^R_V)^4 B_p B_d \big ) 
\end{multline}

This expression contains multiplications of the kinematic functions 
from the top production and decay factorized with the couplings. 
The sum of different terms in the expression reflect the superposition of different states: 
the state with a left-handed operator in the production and in the decay 
of the top quark (first term in Eq.~\ref{me_sq_pr_decay_1}), 
the state with right-handed operator 
in the production and the decay of top quark 
(fourth term in Eq.~\ref{me_sq_pr_decay_1}) and the states with a left-handed 
operator in the production \wtb~vertex and a right-handed operator 
in the decay \wtb~vertex and vice versa (second and third terms 
in Eq.~\ref{me_sq_pr_decay_1}). 

The introduction of a auxiliary vector charged boson ($W_{aux}$) with 
the same properties as the SM W boson but with right-handed (anomalous) 
interaction in the \wtb~vertex (as described in Sec. \ref{Idea}) 
makes possible to simulate the second, third and fourth terms of Eq.~\ref{me_sq_pr_decay_1}.

\begin{figure}[h!]
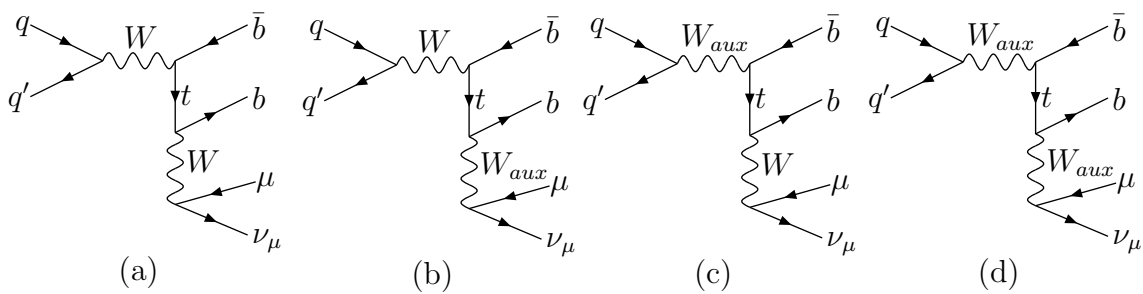

\centering
%\hspace{-0.3in}

\vspace{5mm}
\caption{Diagrams for s-channel single top quark production 
with the additional auxiliary vector field with the same properties as the SM W boson but
 having the right-handed interaction in the \wtb~vertex \label{fig:4diags}
}
\end{figure}

Feynman diagrams help to clarify the terms in Eq.~\ref{me_sq_pr_decay_1}. 
In Fig.~\ref{fig:4diags}(a) the SM diagram for this process is shown. 
Introduction of the new auxiliary field $W_{aux}$ increases the number of diagrams for this process to the four ones; all of them are presented at Fig.~\ref{fig:4diags}~(b), (c) and (d). 

One can combine (include or exclude) some diagrams, it leads to the different operators with different orders of $1/{\Lambda^2}$ which are considered in some or another task. For our present task all four diagrams should be considered; diagrams (b) and (c) correspond to the second and third terms of Eq.~\ref{me_sq_pr_decay_1}
 while the diagram (d) corresponds to the purely right-handed interaction in \wtb~vertices (fourth term in (\ref{me_sq_pr_decay_1})).

\vspace{10mm}
\begin{center} 
\begin{picture}(25,10)(190,70)
\ArrowLine(00.5,99.0)(27.5,85.5) \Text(00.5,99.0)[r]{$q$}
\ArrowLine(27.5,85.5)(00.5,72.0) \Text(00.5,72.0)[r]{$q^{\prime}$}
\Photon(27.5,85.5)(54.5,85.5){3}{3.0} \Text(41.8,90.5)[b]{$W$}
\ArrowLine(81.5,99.0)(54.5,85.5) \Text(83.5,99.0)[l]{$\bar b$}
\ArrowLine(54.5,85.5)(54.5,58.5) \Text(56.8,72.8)[l]{$t$}
\ArrowLine(54.5,58.5)(81.5,72.0) \Text(83.5,72.0)[l]{$b$}
\Photon(54.5,58.5)(54.5,31.5){3}{3.0} \Text(58.8,47.8)[l]{$W$}
\ArrowLine(84.5,40.0)(54.5,31.5) \Text(85.0,40.0)[l]{$\mu$}
\ArrowLine(54.5,31.5)(81.5,20.0) \Text(83.5,18.0)[l]{$\nu_{\mu}$}
\Text(41,0)[b] {(a)}

\end{picture} \

%  diagram # 2
\begin{picture}(25,10)(80,57)
\ArrowLine(00.5,99.0)(27.5,85.5) \Text(00.5,99.0)[r]{$q$}
\ArrowLine(27.5,85.5)(00.5,72.0) \Text(00.5,72.0)[r]{$q^{\prime}$}
\Photon(27.5,85.5)(54.5,85.5){3}{3.0} \Text(41.8,90.5)[b]{$W$}
\ArrowLine(81.5,99.0)(54.5,85.5) \Text(83.5,99.0)[l]{$\bar b$}
\ArrowLine(54.5,85.5)(54.5,58.5) \Text(56.8,72.8)[l]{$t$}
\ArrowLine(54.5,58.5)(81.5,72.0) \Text(83.5,72.0)[l]{$b$}
\Photon(54.5,58.5)(54.5,31.5){3}{3.0} \Text(58.8,47.8)[l]{$W_{aux}$}
\ArrowLine(84.5,40.0)(54.5,31.5) \Text(85.0,40.0)[l]{$\mu$}
\ArrowLine(54.5,31.5)(81.5,20.0) \Text(83.5,18.0)[l]{$\nu_{\mu}$}
\Text(41,0)[b] {(b)}
\end{picture} \

%  diagram # 3
\begin{picture}(25,10)(-25,42)
\ArrowLine(00.5,99.0)(27.5,85.5) \Text(00.5,99.0)[r]{$q$}
\ArrowLine(27.5,85.5)(00.5,72.0) \Text(00.5,72.0)[r]{$q^{\prime}$}
\Photon(27.5,85.5)(54.5,85.5){3}{3.0} \Text(41.8,90.5)[b]{$W_{aux}$}
\ArrowLine(81.5,99.0)(54.5,85.5) \Text(83.5,99.0)[l]{$\bar b$}
\ArrowLine(54.5,85.5)(54.5,58.5) \Text(56.8,72.8)[l]{$t$}
\ArrowLine(54.5,58.5)(81.5,72.0) \Text(83.5,72.0)[l]{$b$}
\Photon(54.5,58.5)(54.5,31.5){3}{3.0} \Text(58.8,47.8)[l]{$W$}
\ArrowLine(84.5,40.0)(54.5,31.5) \Text(85.0,40.0)[l]{$\mu$}
\ArrowLine(54.5,31.5)(81.5,20.0) \Text(83.5,18.0)[l]{$\nu_{\mu}$}
\Text(41,0)[b] {(c)}
\end{picture} \

%  diagram # 4
\begin{picture}(25,10)(-130,27)
\ArrowLine(00.5,99.0)(27.5,85.5) \Text(00.5,99.0)[r]{$q$}
\ArrowLine(27.5,85.5)(00.5,72.0) \Text(00.5,72.0)[r]{$q^{\prime}$}
\Photon(27.5,85.5)(54.5,85.5){3}{3.0} \Text(41.8,90.5)[b]{$W_{aux}$}
\ArrowLine(81.5,99.0)(54.5,85.5) \Text(83.5,99.0)[l]{$\bar b$}
\ArrowLine(54.5,85.5)(54.5,58.5) \Text(56.8,72.8)[l]{$t$}
\ArrowLine(54.5,58.5)(81.5,72.0) \Text(83.5,72.0)[l]{$b$}
\Photon(54.5,58.5)(54.5,31.5){3}{3.0} \Text(58.8,47.8)[l]{$W_{aux}$}
\ArrowLine(84.5,40.0)(54.5,31.5) \Text(85.0,40.0)[l]{$\mu$}
\ArrowLine(54.5,31.5)(81.5,20.0) \Text(83.5,18.0)[l]{$\nu_{\mu}$}
\Text(41,0)[b] {(d)}
\end{picture} \
\end{center}

\vspace{10mm}
%\end{small}

The application for the MC simulation is the following. Only the three sets 
of the event samples are needed for the modeling of the ($f^L_V, f^R_V$) case:
\begin{itemize}
\item the set of events with diagram \ref{fig:4diags}(a) (notation ''1000''),
\item the set of events with pure anomalous  
interaction (right-handed vector operator in the considered case), diagram  \ref{fig:4diags}(d) (notation ''0100''),
\item the set of events with left-handed interaction in the 
top production vertex and right-handed interaction in the top 
decay vertex and vice versa (diagrams  \ref{fig:4diags}(b),(c)) 
(notation ''1100\_artificial''). 
\end{itemize}

The final expression for the simulated event samples combination is:

\begin{equation}
(f^L_V~f^R_V~0~0) = (f^L_V)^4\cdot(1000) + (f^L_V)^2 (f^R_V)^2 \cdot (1100\_artificial) + (f^R_V)^4\cdot(0100),
\label{sigma_lvrv_modelling}
\end{equation}

\begin{figure*}[pH!!!]
\begin{minipage}[b]{.98\linewidth}
\centering
\includegraphics[width=70mm,height=60mm]{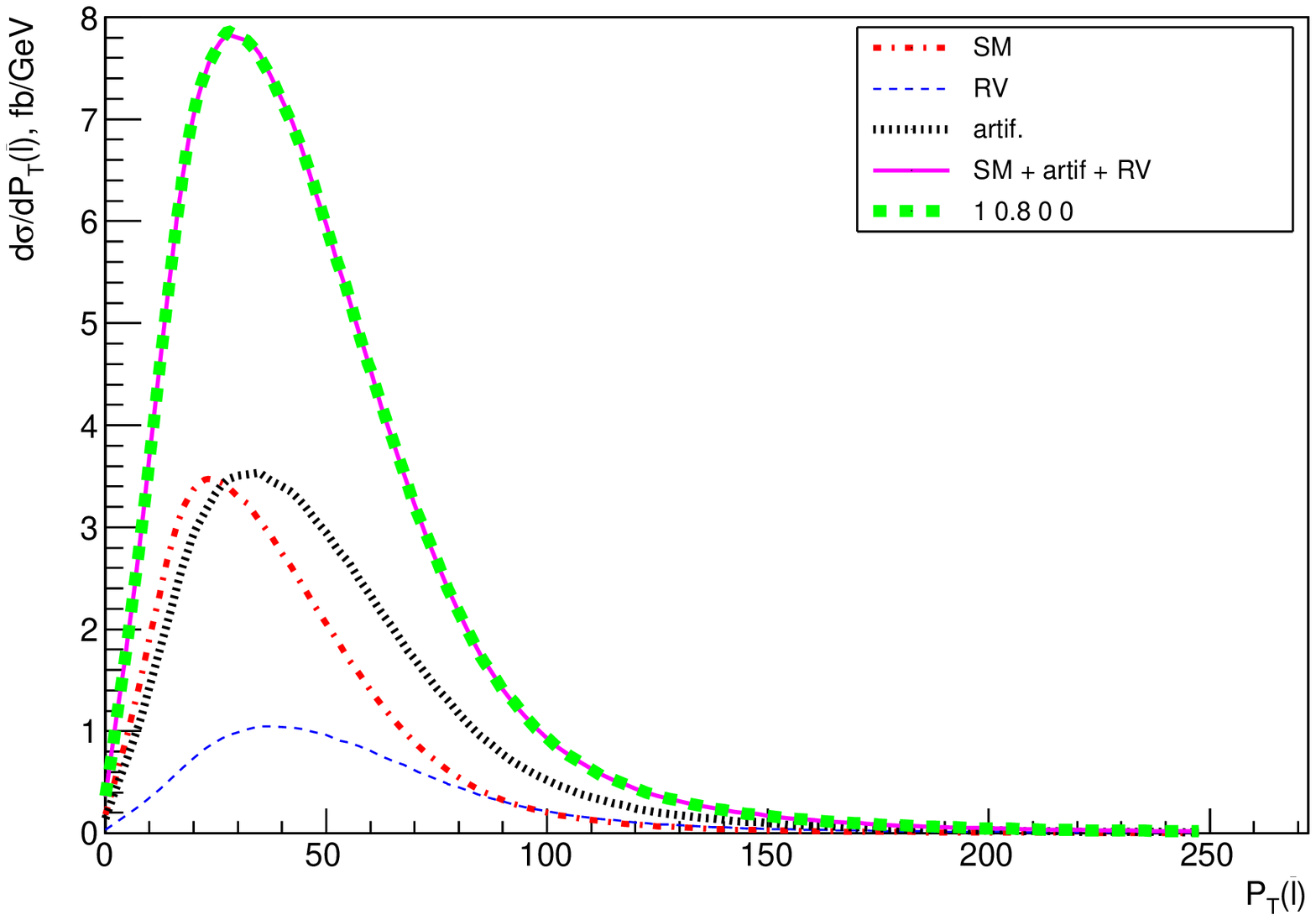}
\includegraphics[width=70mm,height=60mm]{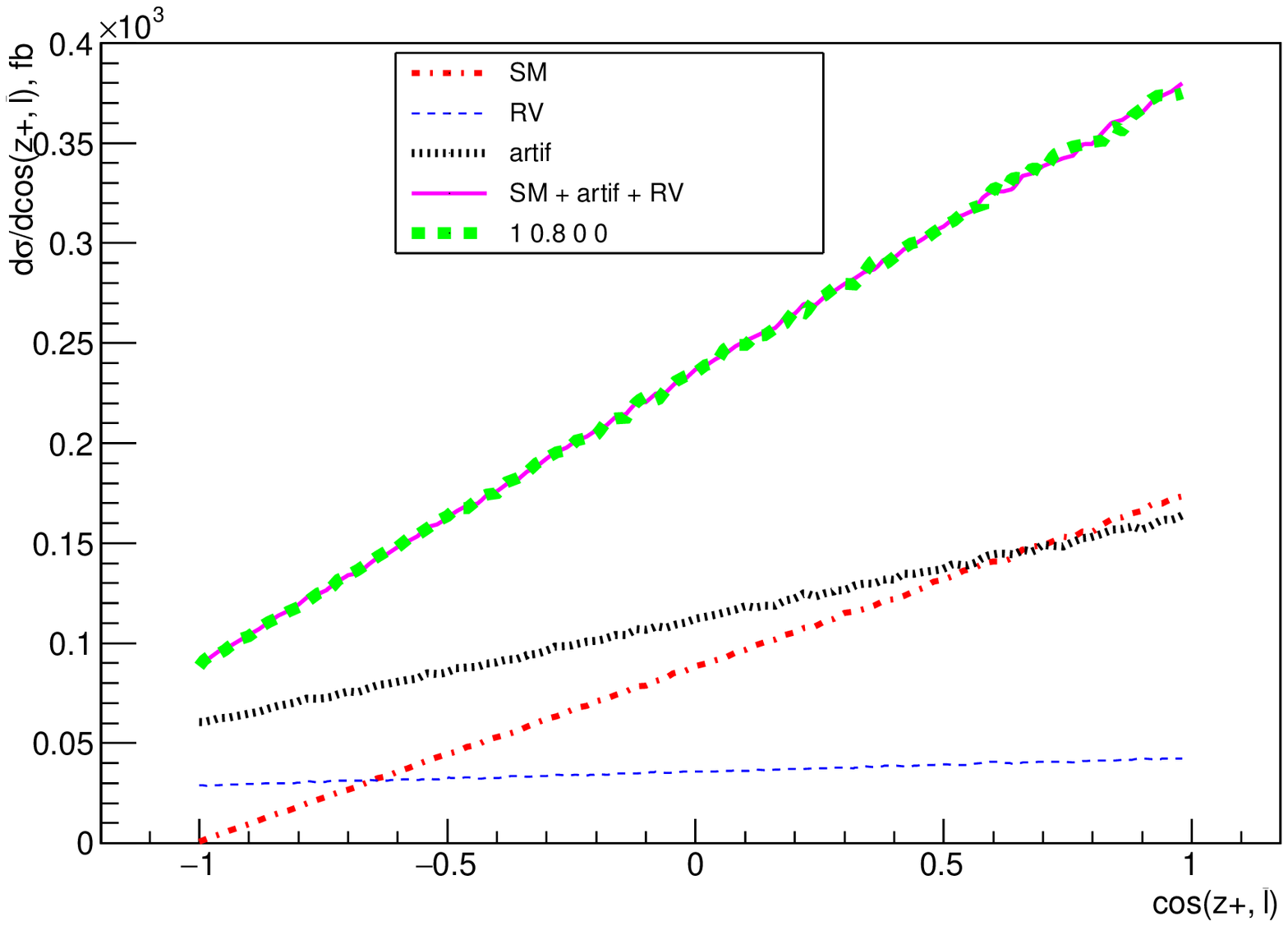}
\Text(-200,-17)[b] {(a)~~~~~~~~~~~~~~~~~~~~~~~~~~~~~~~~~~~~~~~~~~~~~~~~(b)}
\caption {\footnotesize The distributions for the transverse momentum of lepton from top-quark decay in laboratory rest frame (a) and the cosine of angle between lepton from top quark decay and light quark in the top rest frame (b) for the process $pp \rightarrow t (\nu_{\l},\bar {\l}, b) \bar b $ (s-channel single top quark production) for $(f^L_V, f^R_V)$ scenario. \label{LVRV_s_ch_plots} }
\end{minipage}
\end{figure*}

\begin{figure*}[pH!!!]
\begin{minipage}[b]{.98\linewidth}
\centering
\includegraphics[width=70mm,height=60mm]{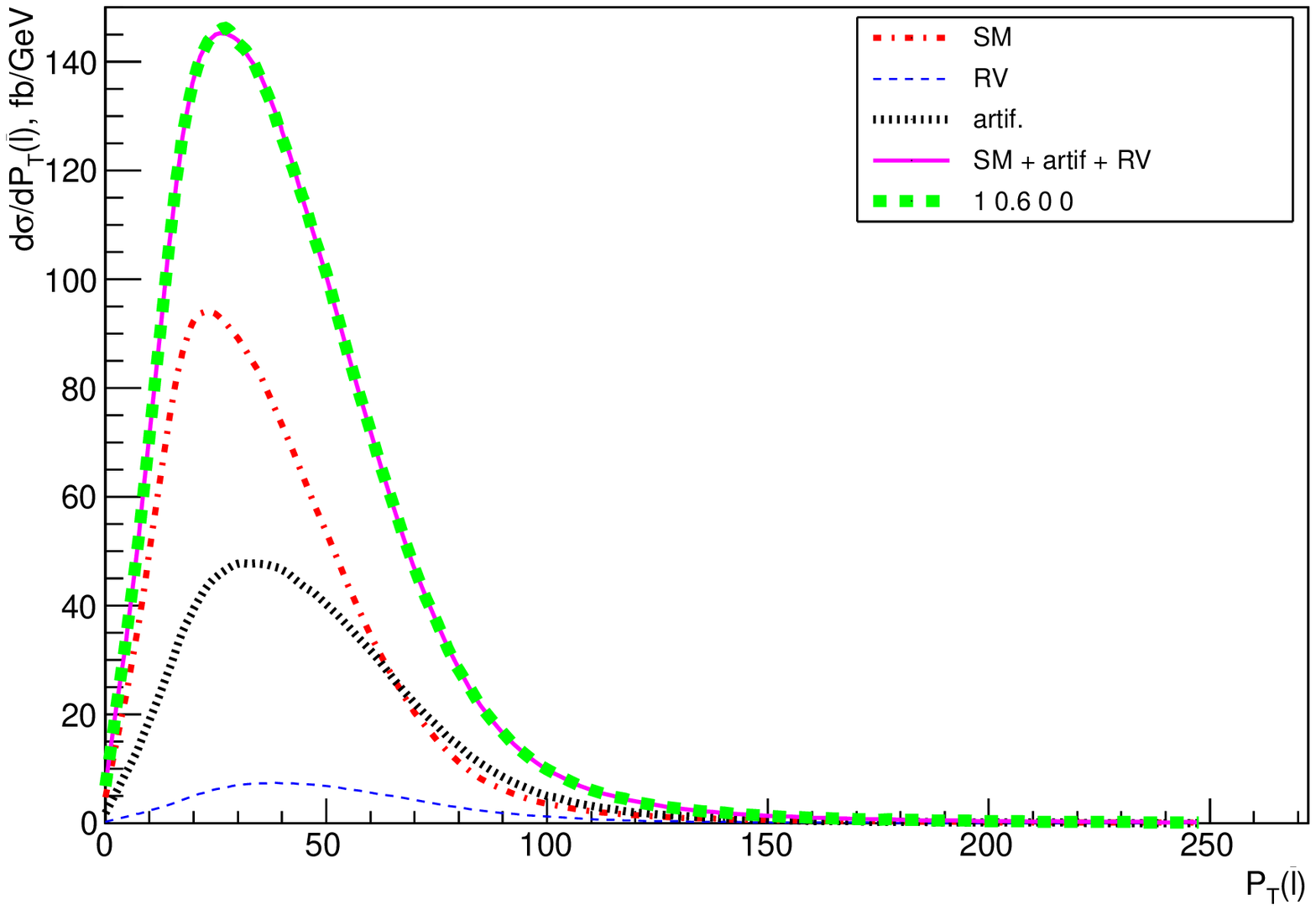}
\includegraphics[width=70mm,height=60mm]{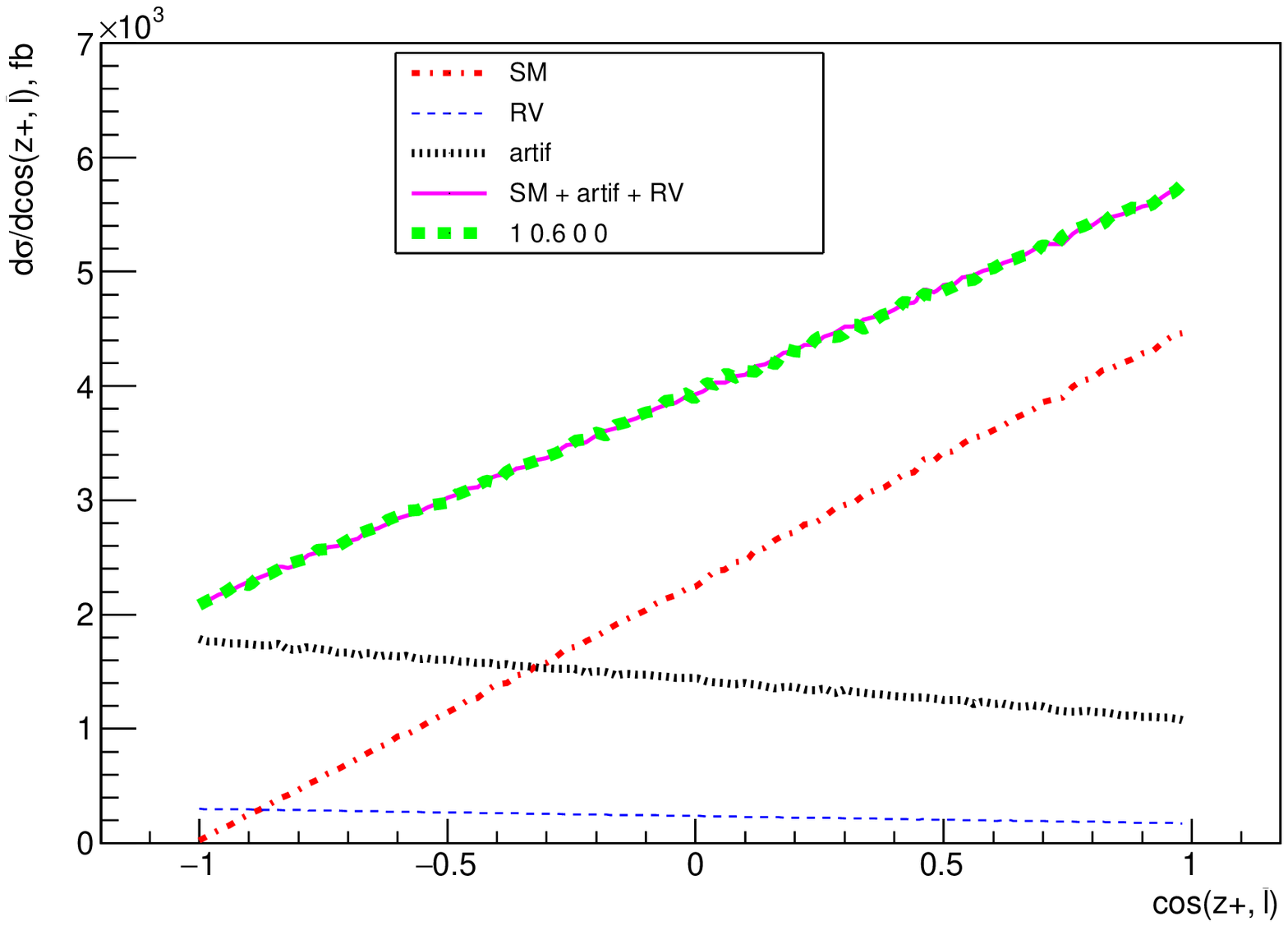}
\Text(-200,-17)[b] {(a)~~~~~~~~~~~~~~~~~~~~~~~~~~~~~~~~~~~~~~~~~~~~~~~~(b)}
\caption {\footnotesize The distributions for the transverse momentum of lepton from top quark decay in laboratory rest frame (a) and the cosine of angle between lepton from the top quark decay and light quark in the top rest frame (b) for the process $pp \rightarrow t (\nu_{\l},\bar {\l}, b) q$ (t-channel single top quark production) for $(f^L_V, f^R_V)$ scenario.. \label{LVRV_t_ch_plots} }
\end{minipage}
\end{figure*}

The ability of the method to simulate different anomalous contributions for the $(f^L_V,  f^R_V)$ scenario
 was tested for the arbitrary values of anomalous
 \wtb~couplings and results are presented in Fig.~\ref{LVRV_s_ch_plots} for s-channel 
single top quark production with $f^L_V=1.0$ and $f^R_V=0.8$ values of couplings and in Fig.~\ref{LVRV_t_ch_plots} for t-channel
 single top quark production for $f^L_V=1.0$ and $f^R_V=0.6$. The notations for $(f^L_V)^4\cdot(1000)$ , $ (f^L_V)^2 (f^R_V)^2 \cdot (1100\_artificial)$ and $(f^R_V)^4\cdot(0100)$ are 
"SM", "artificial" and "RV" in the plots correspondingly.   For these plots, as well as for other plots over the Sec. \ref{subsec:lvrv}-\ref{subsec:lvrt}, the value of top quark total width with the values of constants are equal to  $f^L_V=f^R_V=f^L_T=f^R_T=1$ is used. 
 
 From the plots one can see that, for example, at the Fig.~\ref{LVRV_s_ch_plots}(b) the sum of the curves "SM", "artificial" and "RV" is precisely coinsides with the curve "1 0.8 0 0" which shows the distribution of the cosine of angle between lepton from top quark decay and light quark in the top rest frame in the s-channel single top quark production for the $f^L_V=1.0$ and $f^R_V=0.8$ values of the anomalous \wtb~couplings in \wtb~vertex.  The conclusion is that the presence of left-handed and right-handed vector operators are both in \wtb~vertices is nicely modeled by three sets of events which are correspond to the first ("SM"), second ("artificial") and third ("RV") terms of Eq.~\ref{sigma_lvrv_modelling}.

%=========================================================================
%=========================================================================

%=========================================================================
%=========================================================================

% \newpage
\subsection{$(f^L_V, f^L_T)$ scenario}
\label{subsec:lvlt}
The case with left vector $f^L_V$ and left tensor $f^L_T$ non-zero couplings in \wtb~vertex is similar to the one described 
in Sec. \ref{subsec:lvrv} because the cross term with $f^L_V$ and $f^L_T$ 
multiplication is absent in (\ref{xsection}). The three sets of 
events are needed for accurate simulation of kinematics with all
 possible values of ($f^L_V, f^L_T$) couplings:
\begin{itemize}
\item the "1000'' set of events, diagram from Fig.~\ref{fig:4diags}(a),
\item the "0010" set of events with pure anomalous left tensor
 interaction in the \wtb~vertex (with $f^L_V=f^R_V=f^R_T=0, f^L_T=1$), diagram from Fig.~\ref{fig:4diags}(d),
\item the set of  events with SM interaction 
in the top production vertex and left tenzor coupling in the top 
decay vertex and vice versa (''1010\_artificial'' set of events which is related to the diagrams  at Fig.~\ref{fig:4diags}(b),(c)).
\end{itemize}
The combined expression looks similar to that in Eq.~\ref{sigma_lvrv_modelling}:

\begin{equation}
(f^L_V~0~f^L_T~0) = (f^L_V)^4 \cdot (1000) + (f^L_V)^2 (f^L_T)^2 \cdot (1010\_artificial) +  (f^L_T)^4 \cdot  (0010),
\label{sigma_lvlt_modelling}
\end{equation}

\begin{figure*}[pH!]
\begin{minipage}[b]{.98\linewidth}
\centering
\includegraphics[width=70mm,height=60mm]{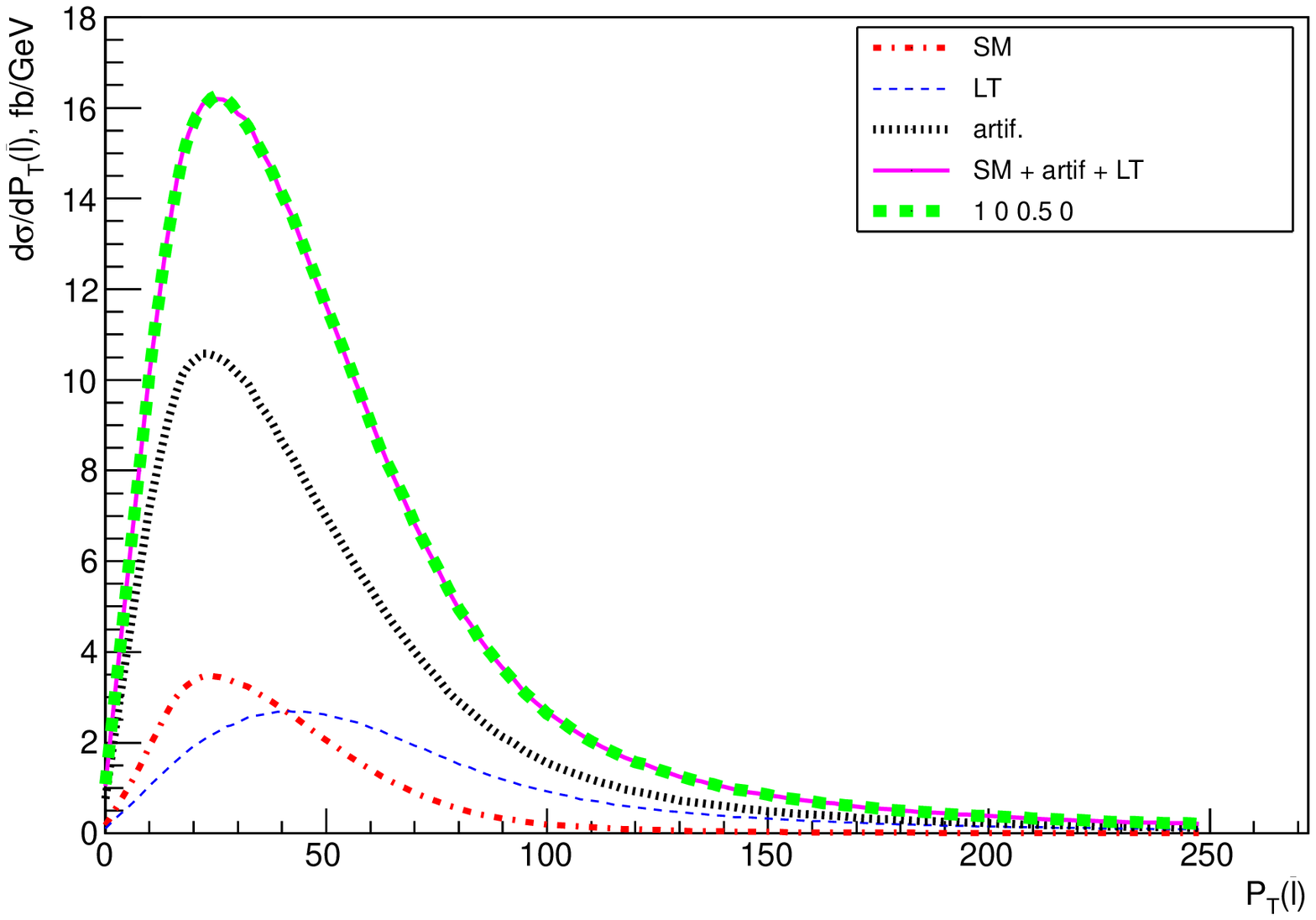}
\includegraphics[width=70mm,height=60mm]{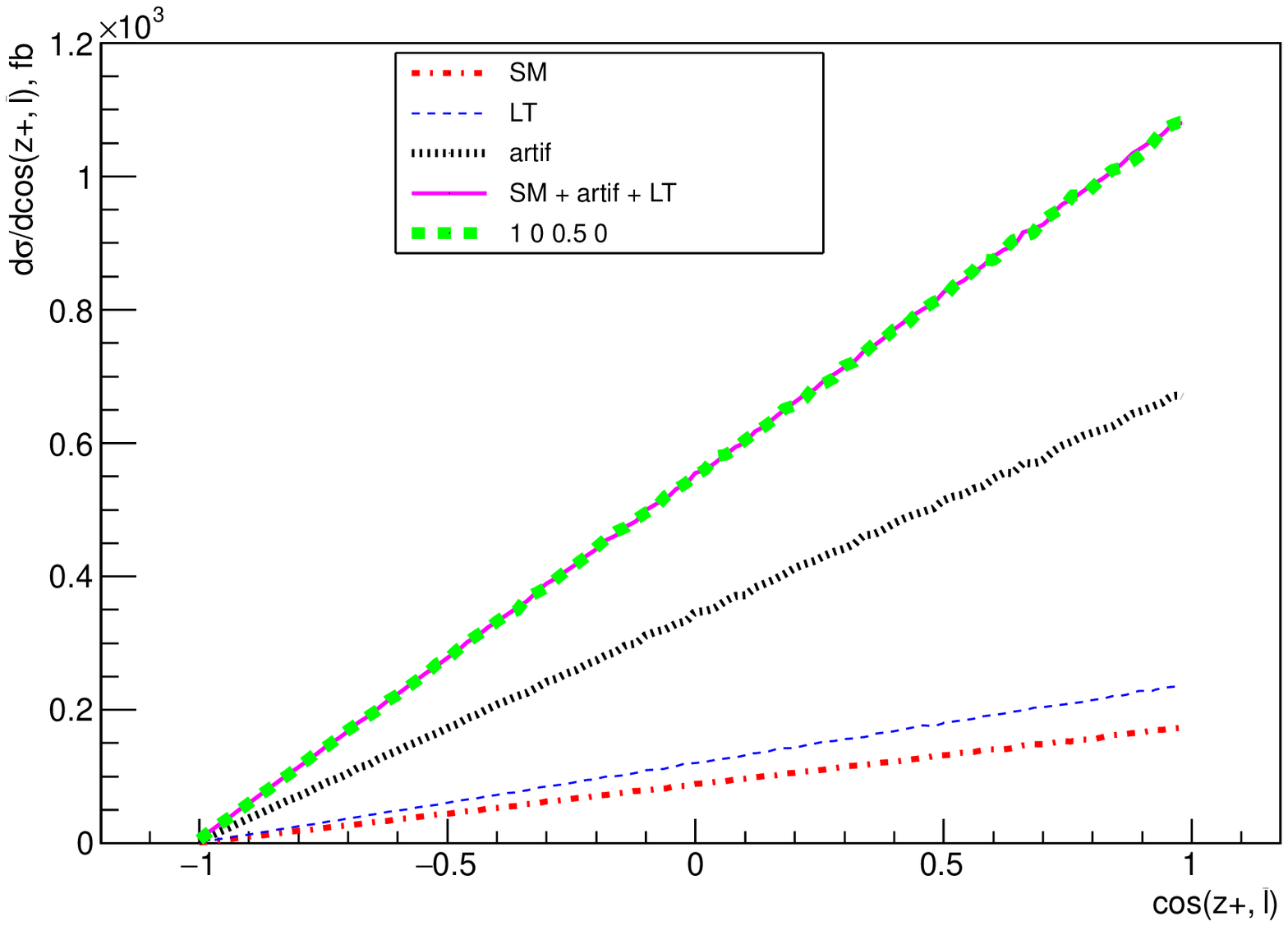}
\Text(-200,-17)[b] {(a)~~~~~~~~~~~~~~~~~~~~~~~~~~~~~~~~~~~~~~~~~~~~~~~~(b)}
\caption {\footnotesize The distributions for the transverse momentum of top-quark in laboratory rest frame (a) and the cosine of angle between lepton from the W-boson decay from top quark and light quark in the top rest frame (b) for the process $pp \rightarrow t (\nu_{\l},\bar {\l}, b) \bar b $ (s-channel single top quark production) for $(f^L_V, f^L_T)$ scenario. \label{LVLT_s_ch_plots} }
\end{minipage}
\end{figure*}

\begin{figure*}[pH!]
\begin{minipage}[b]{.98\linewidth}
\centering
\includegraphics[width=70mm,height=60mm]{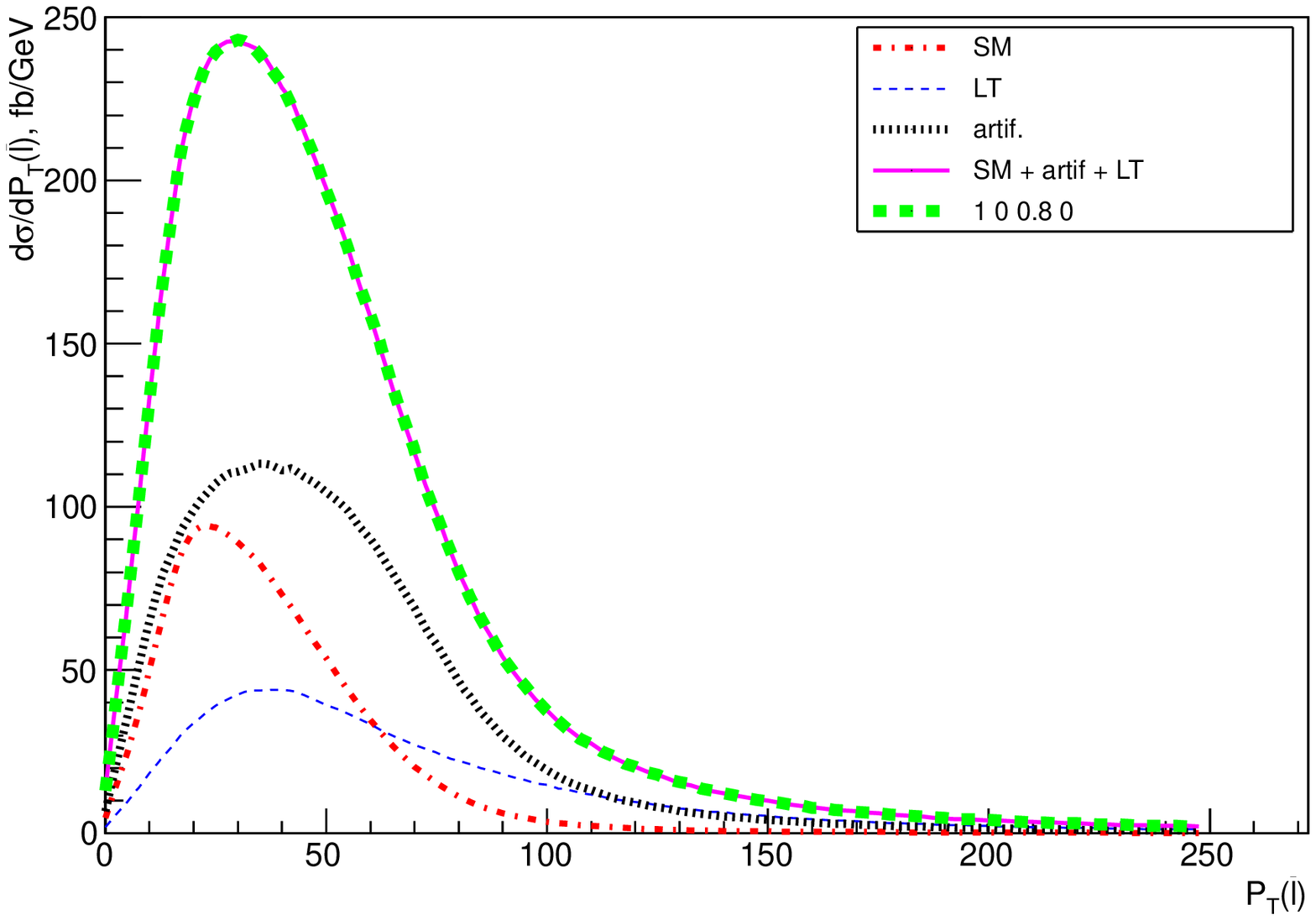}
\includegraphics[width=70mm,height=60mm]{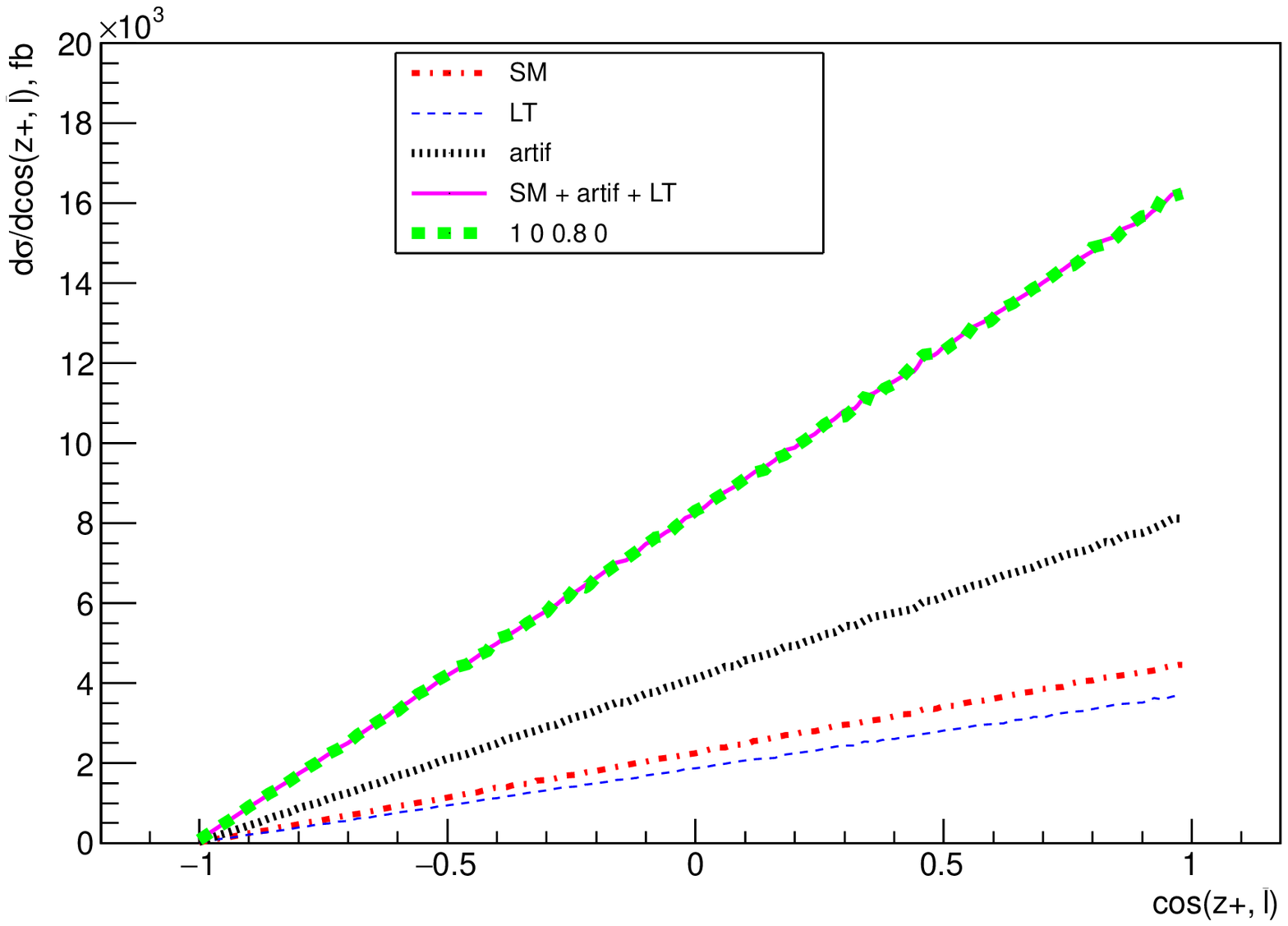}
\Text(-200,-17)[b] {(a)~~~~~~~~~~~~~~~~~~~~~~~~~~~~~~~~~~~~~~~~~~~~~~~~(b)}
\caption {\footnotesize The distributions for the transverse momentum of top-quark in laboratory rest frame (a) and the cosine of angle between lepton from the W-boson decay from top quark and light quark in the top rest frame (b) for the process $pp \rightarrow t (\nu_{\l},\bar {\l}, b) q$ (t-channel single top quark production) for $(f^L_V, f^L_T)$ scenario..  \label{LVLT_t_ch_plots} }
\end{minipage}
\end{figure*}

Figures~\ref{LVLT_s_ch_plots} and \ref{LVLT_t_ch_plots} demonstrate 
correct reproduction of the case with $f^L_V = 1, f^L_T=0.5$ and $f^L_V = 1, f^L_T=0.8$ values of 
the couplings in the \wtb~vertex by means of the combination "SM", "artificial" and "LT" (first, second and third terms in (\ref{sigma_lvlt_modelling})) sets of events.

%=========================================================================
%=========================================================================

%=========================================================================
%=========================================================================
% \newpage
\subsection{$(f^L_V, f^R_T)$ scenario}
\label{subsec:lvrt}

The scenario with $f^L_V$ and $f^R_T$ couplings  in \wtb~vertex is the more complicate than the ones described in subsections  
\ref{subsec:lvrv} and \ref{subsec:lvlt} due to the presence 
of the ($f^L_V \cdot f^R_T$) cross term in Eq.~\ref{xsection}.

In this scenario five kinematic terms with different 
powers of constants $f^L_V$ and $f^R_T$ arise in the squared matrix element:

\begin{equation}
\label{me_sq_lvrt_pr_decay}
|M|^{2}_{tot} \sim \sum^{4}_{i=0} k_{i} \cdot P_{4-i} D_i 
\end{equation}

where $P,D$ are some kinematic functions of the production and decay 
of top quarks and 
$k_{i}=(f^L_V)^{4-i} (f^R_T)^{i} $ (upper $i$ is power, not an indices).

The idea is to combine event samples which related to the same powers of constants in (\ref{me_sq_lvrt_pr_decay}). At the computational level it means the selection of the squared diagrams  for the process of single top quark production and the subsequent decay with the SM W-boson and $W_{aux}$; the last one has the right tenzor coupling with top and bottom quarks in the \wtb~vertex and the SM-like behaviour in all other vertices. E.g. the term $k_{1}=(f^L_V)^{3} (f^R_T)^{1}\cdot P_3 D_1$ from Eq. \ref{me_sq_lvrt_pr_decay}
corresponds to the set of squared diagrams which have three W-bosons and one $W_{aux}$ boson.

One should note that some terms in  Eq.~\ref{me_sq_lvrt_pr_decay} with the 
odd powers of anomalous couplings  have negative contribution to the total cross section and 
one needs to keep it in mind while generating the sets of events. 

The following mimimal set of event samples is needed for correct reproduction of the case with arbitrary values of $f^L_V$, $f^R_T$ couplings are in the \wtb~vertex:

\begin{itemize}
\item the "1000'' set of events; it corresponds to the squared diagram if one squares the diagram from  Fig.~\ref{fig:4diags}(a) (notation "1000"),
\item the "0001" set of events  with pure anomalous right tensor
 interaction in the \wtb~vertex (with $f^L_V=f^R_V=f^L_T=0, f^R_T=1$);  it corresponds to the squared diagram if one squares the diagram from  Fig.~\ref{fig:4diags}(d) (notation "0001"),
\item the 3 additional sets of  events; they correspond to the all other squared diagrams if one squares the whole set of diagrams from  \ref{fig:4diags}. All of these samples are combined in correspondence with the powers of couplings $f^L_V$ and $f^R_T$ in Eq. \ref{me_sq_lvrt_pr_decay}. 
\end{itemize}

\begin{figure*}[pH!!!]
\begin{minipage}[b]{.98\linewidth}
\centering
\includegraphics[width=70mm,height=60mm]{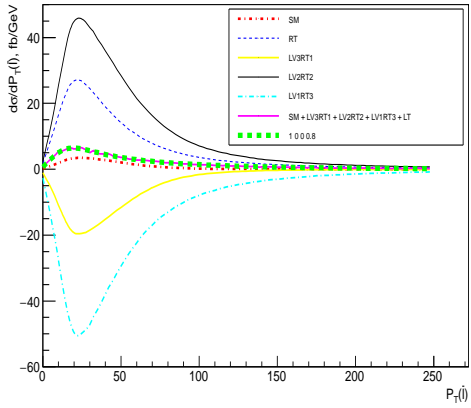}
\includegraphics[width=70mm,height=60mm]{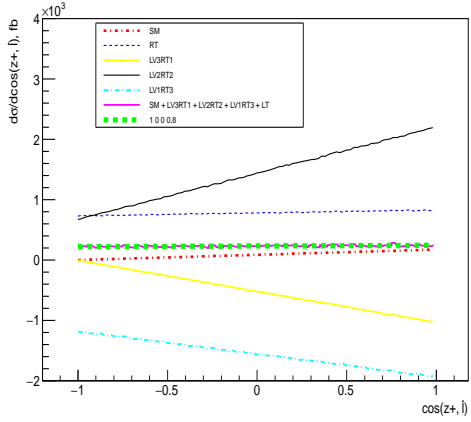}
\Text(-200,-17)[b] {(a)~~~~~~~~~~~~~~~~~~~~~~~~~~~~~~~~~~~~~~~~~~~~~~~~(b)}
\caption {\footnotesize The distributions for the transverse momentum of lepton from top quark decay in laboratory rest frame (a) and the cosine of angle between lepton from top quark decay and light quark in the top rest frame (b) for the process $pp \rightarrow t (\nu_{\l},\bar {\l}, b) \bar b $ (s-channel single top quark production) for $(f^L_V, f^R_T)$ scenario. \label{LVRT_s_ch_plots} }
\end{minipage}
\end{figure*}

\begin{figure*}[pH!!!]
\begin{minipage}[b]{.98\linewidth}
\centering
\includegraphics[width=70mm,height=60mm]{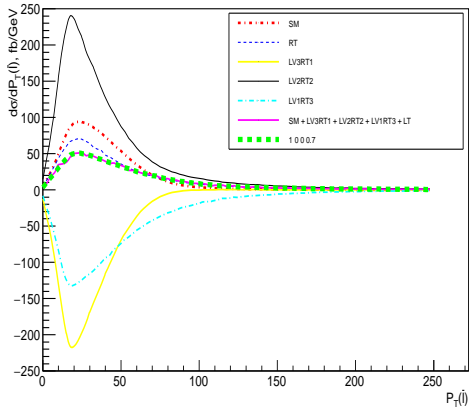}
\includegraphics[width=70mm,height=60mm]{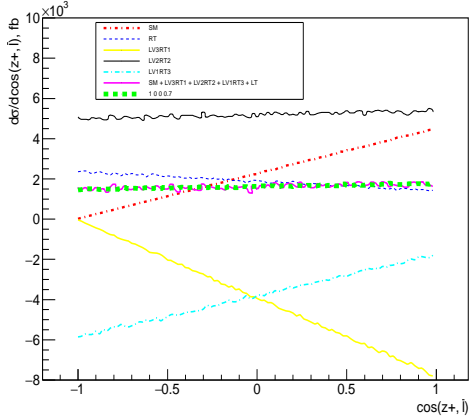}
\Text(-200,-17)[b] {(a)~~~~~~~~~~~~~~~~~~~~~~~~~~~~~~~~~~~~~~~~~~~~~~~~(b)}
\caption {\footnotesize The distributions for the transverse momentum of top quark in laboratory rest frame (a) and the cosine of angle between lepton from the W-boson decay from top quark and light quark in the top rest frame (b) for the process $pp \rightarrow t (\nu_{\l},\bar {\l}, b) q$ (t-channel single top quark production) for $(f^L_V, f^R_T)$ scenario..  \label{LVRT_t_ch_plots} }
\end{minipage}
\end{figure*}

For the illustration of the method the case with $f^L_V=1.0$ 
and $f^R_T=0.8$ and $f^R_T=0.7$ values of the couplings in \wtb~vertex is simulated 
with the help of the described mimimal set of event samples. The results are shown in Figs.~\ref{LVRT_s_ch_plots} 
and~\ref{LVRT_t_ch_plots}. For example, in \ref{LVRT_s_ch_plots}(a) the curve "1 0 0 0.8" shows the distribution of the transverse momenta of the lepton from the top quark decay for the case with $f^L_V=1.0$ and $f^R_T=0.8$ values of the anomalous couplings in \wtb~vertex. The curve "SM" shows  the same  distribution following from $(f^L_V)^4\cdot(1000)$ set of events and 
the curve "RT" shows  the same  distribution following from  $(f^R_T)^4\cdot(0001)$ set of events.
The curves "LV3RT1", "LV2RT2"...  show the distributions following from the event sets which correpond to the parts of (\ref{me_sq_lvrt_pr_decay}) and the squared diagrams with SM W-boson and $W_{aux}$,  multiplied by the factors $(f^L_V)^3(f^R_T)$, $(f^L_V)^2(f^R_T)^2$ etc. One can see the precise convergence between "1 0 0 0.8" curve and the sum of the five event sets which are described above.
 
%=========================================================================
%=========================================================================

%=========================================================================
%=========================================================================

Since the cross terms between
 $f_{\rm T}^{\rm L}$ and $f_{\rm T}^{\rm R}$ or  $f_{\rm V}^{\rm R}$ 
and $f_{\rm T}^{\rm R}$ couplings are suppressed, it is possible 
to use the event samples described above to simulate kinematics with 
three-dimensional variation of the 
$f_{\rm V}^{\rm L}$, $f_{\rm T}^{\rm L}$, $f_{\rm T}^{\rm R}$
 or $f_{\rm V}^{\rm L}$, $f_{\rm V}^{\rm R}$, $f_{\rm T}^{\rm R}$ couplings.
 For the simulation of kinematics with all four couplings additional samples 
are needed.

\subsection{Application of the method to the experiment}
\label{subsec:experiment}

One needs to say a few words concerning the application of this method to the experimental analyses of the single top quark production processes. 

First assumption that has been accepted is the massless b-quark. This assumption leads to  the simple expression of the cross section dependence on the anomalous couplings (\ref{xsection}) with the following simple conception statements, i.e. three considered scenarios etc. The existed non-zero b-quark mass leads to the cross terms in (\ref{xsection}) not only for  $(f^L_V, f^R_T)$ and $(f^R_V, f^L_T)$ couplings but for all the couplings. However these terms are suppressed by the factor of $(\frac{m_b}{m_{top}})^2$ which is sufficiently tiny for the experiment.

The second assumption is more important for the experimental application of the method. The minimal set of event samples, e.g. for the $(f^L_V, f^L_T)$ scenario, consists of three set of events: "1000", "1010\_artificial" and "0010". The same top quark width, calculated with all four anomalous couplings are equal to 1, has been used for all the plots and for all sets of events. However the presence of anomalous operators significantly changes the top quark total width. To take this effect into account for modeling of the anomalous events one needs, following the narrow width approximation, to apply the factors $\frac{w_s}{w_{tot}}$ where $w_s$ is the value of total top quark width for each set of events from the minimal basis and $w_{tot}$ is the total top quark width with non-zero anomalous couplings~\cite{Mohammadi Najafabadi:2006um}. The expression (\ref{sigma_lvlt_modelling}) with the adjusted factors is the following:
\begin{equation}
(f^L_V~0~f^L_T~0) = m \cdot (1000) + n \cdot (1010\_artificial) +  k \cdot  (0010),
\label{sigma_lvlt_modelling_corrections}
\end{equation}
where  $m = \left( f^L_V \right)^{4}  w_{1000}/w_{tot}^{f^L_V,f^L_T}$,  
$n = \left( f^L_V \right)^{2}\left( f^L_T \right)^{2}  w_{1010\_artificial}/w_{tot}^{f^L_V,f^L_T}$ and \\
$k = \left( f^L_T \right)^{4} w_{0010}/w_{tot}^{f^L_V, f^L_T}$. 

\begin{figure*}[!h!]
\begin{minipage}[b]{.98\linewidth}
\centering
\includegraphics[width=70mm,height=60mm]{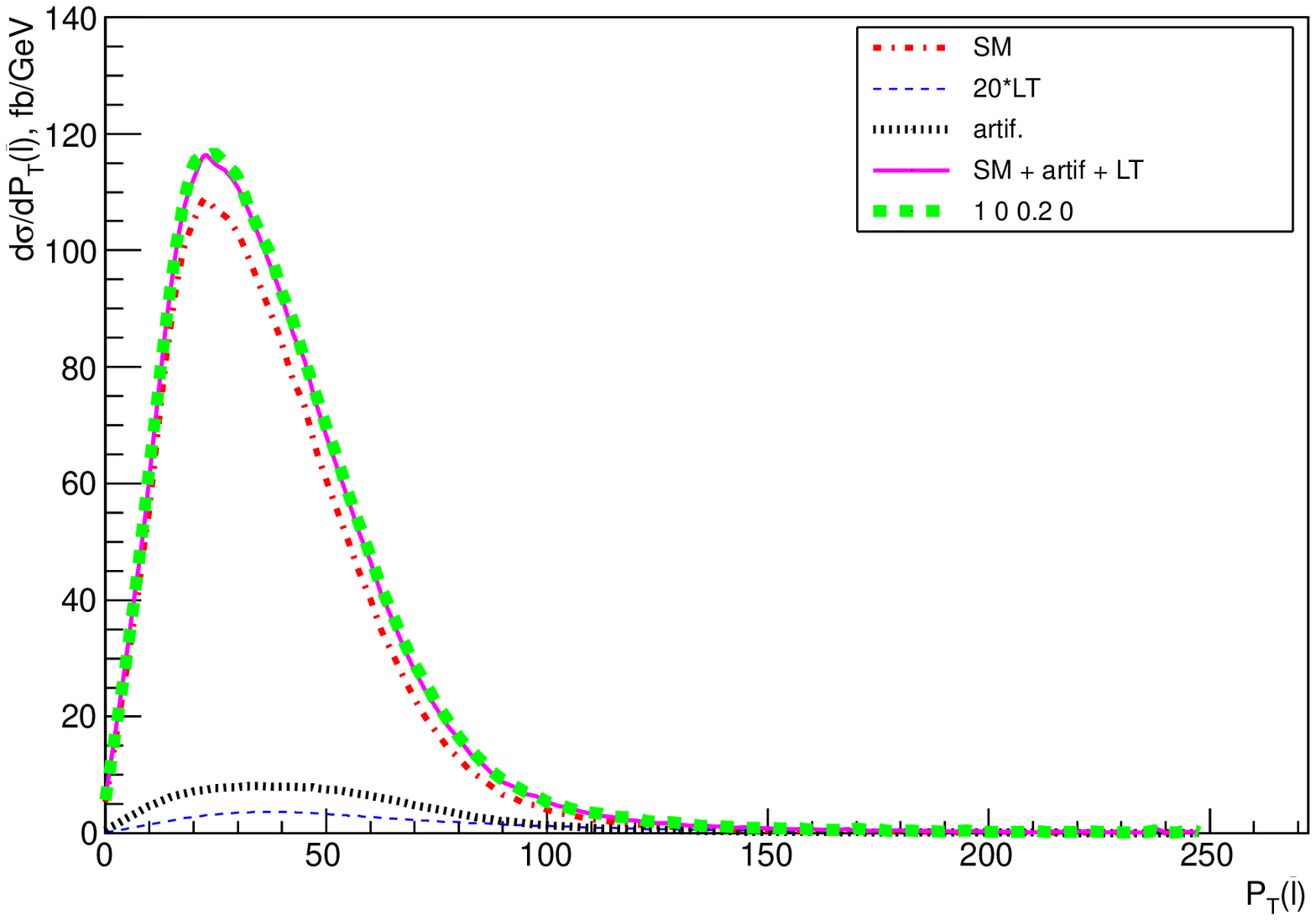}
\includegraphics[width=70mm,height=60mm]{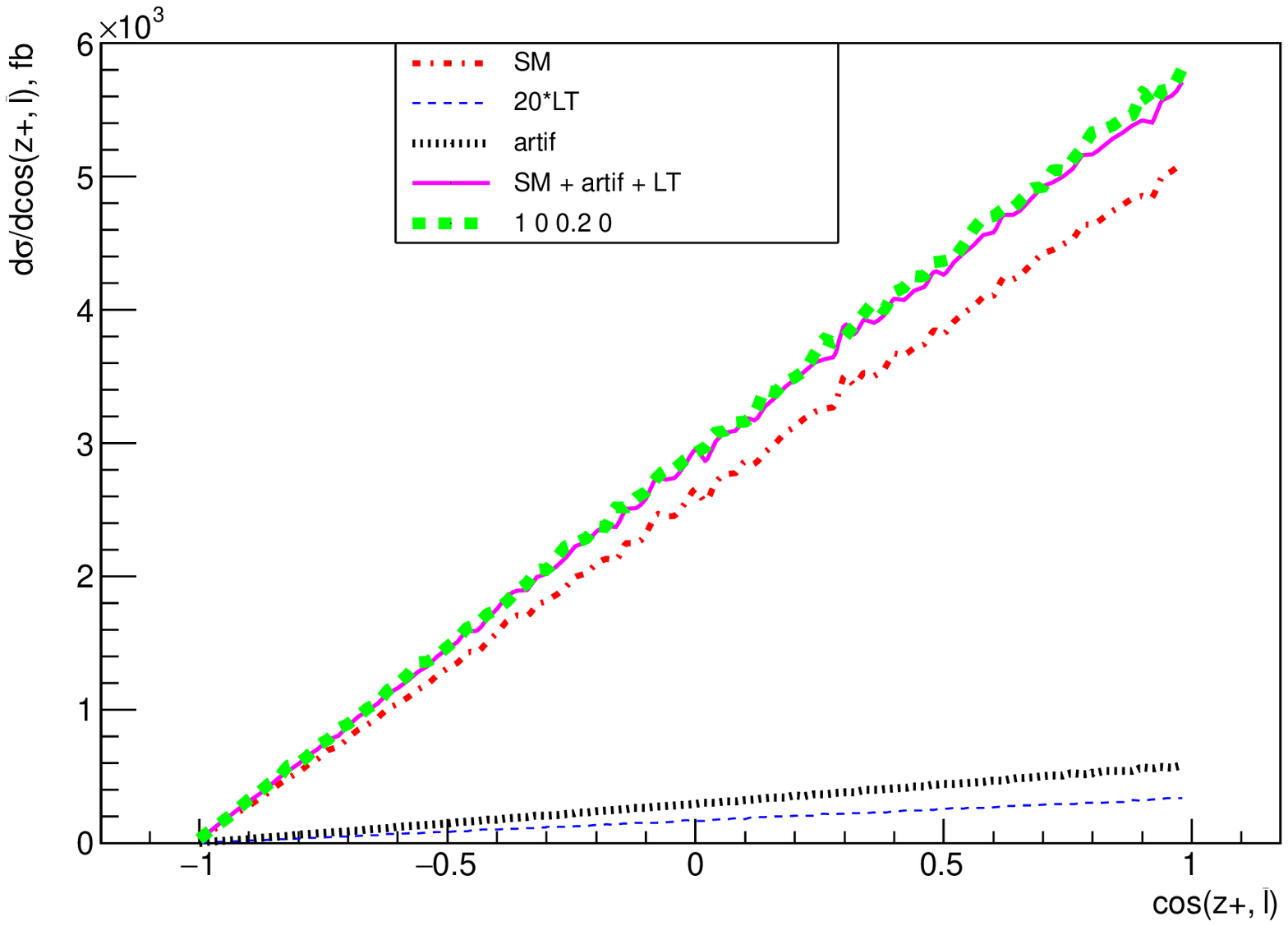}
\Text(-200,-17)[b] {(a)~~~~~~~~~~~~~~~~~~~~~~~~~~~~~~~~~~~~~~~~~~~~~~~~(b)}
\caption {\footnotesize The distributions for the transverse momentum of top-quark in laboratory rest frame (a) and the cosine of angle between lepton from the W-boson decay from top quark and light quark in the top rest frame (b) for the process $pp \rightarrow t (\nu_{\l},\bar {\l}, b) q$ (t-channel single top quark production) for $(f^L_V, f^L_T)$ scenario without the assumption of massless b-quark and with the adjusted factors which are related to different values of top quark widths with anomalous couplings are present.   \label{LVLT_corr_t_ch_plots} }
\end{minipage}
\end{figure*}

Figure~\ref{LVLT_corr_t_ch_plots} demonstrates that the factors in (\ref{sigma_lvlt_modelling_corrections}) should be included to reproduce correctly the total result by the sum of the individual contributions. The formula (\ref{sigma_lvlt_modelling_corrections}) shows how to get the event set with arbitrary values of the anomalous couplings from the basic event sets with anomalous couplings taken to be one or zero. In this example in the Fig.~\ref{LVLT_corr_t_ch_plots} non-zero b-quark mass was taken and as expected the influence of non-zero b-quark mass is negligible.

The event sets of the single top quark production including anomalous Wtb coupling are prepared as described above for the LHC energies and posted recently into the Monte-Carlo simulated event database~\cite{Belov:2007qg}.

%=========================================================================
%=========================================================================
One should also notice, since the $W_{aux}$ interacts with other particles by electroweak 
forces the introduction of this particle does not affect the NLO QCD 
corrections, therefore the method is applicable at the NLO level as well.
\section{Conclusion}
\label{sec:Conclusion}

In this paper we present a new method of modeling the events with anomalous fermion-boson couplings which is based on introducing some auxiliary vector fields 
in addition to the SM gauge field in the unitary gauge. The auxiliary fields have the same masses and all the couplings to fermions as the SM gauge field except the couplings to the fermion with anomalous interaction. The coupling of the auxiliary field to that fermion is the anomalous coupling. The method allows to perform simulations in two different approaches keeping only the linear order contributions or keeping higher order contributions in anomalous couplings. The first approuch is motivated by the effective field theory (EFT) in which the only leading  $1/{\Lambda}^2$ contributions are taken into account. In the second approach higher orders in $1/{\Lambda}^2$ are also taken into account as appeared in direct matrix element computations. Since each of the anomalous coupling is associated with corresponding auxiliary field it is very easy to keep only needed contribution by removing all not needed diagrams from the amlitude or squared diagrams from the matrix element squared. In fact, it is very instructive to use both approches simultaneusly since a comparison of results in two cases allows to understand a region of appicability (EFT) in the anomalous parameter space. 

The method also allows to simplify significantly realistic analyses by generating  
only the minimum number of the event samples with the unity values of the anomalous couplings and very easy to be implemented in computing code, as was done in this study using CompHEP. The proposed method works for arbitary widths of the fermion resonanses. However the simple formula to rescale contributions from different sets of events (such as Eq.~\ref{sigma_lvlt_modelling_corrections}) works only in the narrow width approximation.

Practical use of proposed method is demonstrated in an example of the single top quark production  processes with anomalous \wtb~couplings.  In our demonstration we focused on more difficult for the analysis approach computing Feynman diagrams with non-linear behaviour of the anomalous couplings. In this case the terms with higher dimensions on $1/{\Lambda}^2$ arise due to the multiplication of the production and decay parts of the processes both depending on anomalous couplings and the presence of the total top quark width in the denominator. One should stress that if one considers only the leading terms of the order of $1/{\Lambda}^2$ one needs not only to keep leading terms in numerator of diagrams but also to expand the total top quark width in the denominator and to take into account the terms with the dimension of $1/{\Lambda}^2$ in the overall expansion. 
\section{Acknownlegements}
\label{sec:acknownlegements}

The authors are grateful to R.Schwienhorst and H. Prosper as well as many colleagues drom \dzero and \cms single top groups for useful discussions. The work of M. Perfilov was partially supported by Russian Foundation "Dynasty".

\end{document}